\RequirePackage{lineno}
\documentclass[twocolumn,showpacs,aps,prl,superscriptaddress]{revtex4}
\usepackage{xspace}
\usepackage{graphicx}
\usepackage{amsmath}
\usepackage{amssymb}
\usepackage{epsfig}

\def\asy         {\ensuremath{A_Q}\xspace }

\def\gpiz        {\ensuremath{(\geq \! 0\piz)}\xspace}

\def\taupks      {\ensuremath{\taum\rightarrow\pim\KS\,\nut}\xspace }

\def\taupksgpiz   {\ensuremath{\taum\rightarrow\pim\KS\,\gpiz\,\nut}\xspace }

\def\taupksthrpiz   {\ensuremath{\taum\rightarrow\pim\,\KS \, 3\piz\nut }\xspace }

\def\taupksb     {\ensuremath{\taup\rightarrow\pip\KS\ \nutb}\xspace }

\def\taukks      {\ensuremath{\taum\rightarrow\Km\KS\ \nut}\xspace }

\def\taukksgpiz   {\ensuremath{\taum\rightarrow\Km\KS\gpiz\nut}\xspace }

\def\taupkzkz{\ensuremath{\taum\rightarrow\pim\Kz\Kzb \nut}\xspace }

\def\totsyse    {0.13} 

\def\ndatpe     {\ensuremath{99842}}
\def\ndatne     {\ensuremath{99222}}
\def\ndate      {\ensuremath{199064}}

\def\nbkgpe     {1393}
\def\nbkgne     {1401}
\def\ebkgpe     {79} 
\def\ebkgne     {74} 

\def\deterre    {0.12} 
\def\bkgerre    {0.05}

\def\corrtaue   {0.81}
\def\ecorrtaue	{0.03}
\def\corrkse    {0.9}
\def\ecorrkse	{0.4}

\def\corrtaum	{0.49}
\def\ecorrtaum	{0.03}
\def\corrksm	{1.0}
\def\ecorrksm	{0.4}

\def\totsysm    {\ensuremath{0.10}}
 
\def\ndatm      {\ensuremath{140602}}
\def\ndatpm     {\ensuremath{70369}}
\def\ndatnm     {\ensuremath{70233}}

\def\nbkgpm     {1120}
\def\nbkgnm     {1055}
\def\ebkgpm     {65} 
\def\ebkgnm     {74} 

\def\deterrm    {\ensuremath{0.08}}
\def\bkgerrm    {\ensuremath{0.06}}

\def\aeraw{(-0.32\pm 0.23)\% }
\def\amraw{(-0.05\pm 0.27)\% }
\def\aed{(-0.39\pm 0.23\pm 0.13)\% }
\def\amd{(-0.12\pm 0.27\pm 0.10)\% }
\def\amean{(-0.27\pm 0.18\pm 0.08)\% }
\def\afinal{(-0.36\pm 0.23\pm 0.11)\% }
\def\ad{(0.07\pm 0.01)\% }
\def\nstd{ 2.8}
\def\kzerrnew{0.01}

\include{babarsym}

\def\tks{t / \tau_{\KS}}

\begin{document}

\begin{flushleft}

\mbox{\normalsize {BABAR-PUB-11/009} }
\newline
\mbox{\normalsize {SLAC-PUB-14556} }
\end{flushleft}

%
\author{J.~P.~Lees}
\author{V.~Poireau}
\author{V.~Tisserand}
\affiliation{Laboratoire d'Annecy-le-Vieux de Physique des Particules (LAPP), Universit\'e de Savoie, CNRS/IN2P3,  F-74941 Annecy-Le-Vieux, France}
\author{J.~Garra~Tico}
\author{E.~Grauges}
\affiliation{Universitat de Barcelona, Facultat de Fisica, Departament ECM, E-08028 Barcelona, Spain }
\author{M.~Martinelli$^{ab}$}
\author{D.~A.~Milanes$^{a}$}
\author{A.~Palano$^{ab}$ }
\author{M.~Pappagallo$^{ab}$ }
\affiliation{INFN Sezione di Bari$^{a}$; Dipartimento di Fisica, Universit\`a di Bari$^{b}$, I-70126 Bari, Italy }
\author{G.~Eigen}
\author{B.~Stugu}
\affiliation{University of Bergen, Institute of Physics, N-5007 Bergen, Norway }
% ...
\collaboration{The \babar\ Collaboration}
\noaffiliation

\title{Search for \boldmath{\CP} Violation in the Decay \boldmath{$\taum\to\pim\KS\left(\geq 0\piz\right)\nut$}}

\date{January 16, 2012}

\vspace{1pc}
\begin{abstract}
\begin{center}
\large \bf Abstract
\end{center}
We report a search for \CP violation in the decay \taupksgpiz using 
a dataset of 437 million $\tau$ lepton pairs, corresponding to an integrated 
luminosity of $476\,\invfb$, collected with the \babar\ detector at the 
\pep2\ asymmetric energy \epem storage rings.  
The \CP-violating decay-rate asymmetry is determined to be $\afinal$
approximately $\nstd$  standard deviations from the Standard Model prediction of 
$(0.36 \pm 0.01)\%$.
\vspace{1pc}
\end{abstract}


\maketitle

\linenumbers


\CP violation has been observed only in the $K$ and $B$ meson systems.  
However, Bigi and Sanda \cite{bs} predict that, in the Standard Model (SM), 
the decay of the $\tau$ lepton to final states containing a \KS meson will 
also have a non-zero decay-rate asymmetry due to \CP violation in the kaon sector.  
The decay-rate asymmetry 

\begin{equation*}
{\asy} = \frac{\Gamma\left({\taupksb}\right) - \Gamma\left({\taupks}\right)}
              {\Gamma\left({\taupksb}\right) + \Gamma\left({\taupks}\right)}
\label{eq:asy}
\end{equation*}

\noindent
is predicted to be $\left(0.33\;\pm0.01\right)\%$  for decay times comparable
to the lifetime $\tau_{\KS}$ of the \KS meson.
In a recent paper, Grossman and Nir  \cite{grossman} point out that 
Sanda and Bigi did not include the interference between the amplitudes 
of intermediate \KS and \KL which is as important as the pure \KS amplitude.
Therefore the decay-rate asymmetry depends on the reconstruction efficiency 
as a function of the $\KS \rightarrow \pip\pim$ decay time.
If the selection is fully efficient for decay times that are long compared
with the \KS lifetime, then the predicted decay-rate asymmetry  
is almost unchanged relative to the prediction of Bigi and Sanda \cite{bs},
due to a sign error \cite{grossman}.

If the measured decay-rate asymmetry shows a significant deviation from the 
SM value then this could be evidence for new physics.  
No evidence for \CP violation has been found in related studies by \babar\
and Belle
in $\Dp \rightarrow \KS \pip$ decays  \cite{cenci, belle:d}, by the Belle 
collaboration in a study of the angular distribution of the decay 
products in \taupks decays \cite{belle:cpv}, 
or by the CLEO collaboration \cite{cleo:cpv}.

This paper presents a measurement of \asy using \taupksgpiz 
and charge conjugate decays.  
The SM asymmetry is identical for decays with any number of \piz mesons.
If there is an asymmetry due to new-physics dynamics, then the impact 
of including modes with one or more \piz mesons may be different.

The analysis uses data recorded by the \babar\ detector at the \pep2\ 
asymmetric-energy \epem\ collider, operated at center-of-mass (CM) 
energies of 10.58\gev and 10.54\gev at the SLAC National Accelerator Laboratory.
The \babar\ detector is described in detail in Ref.~\cite{detector}.  
In particular, charged kaons and pions are differentiated 
by ionization ($dE/dx$) measurements in the silicon vertex detector
and the drift chamber
in combination with an internally reflecting Cherenkov detector, 
with identification efficiency greater than 90\% for pions 
and kaons with momenta above 1.5\gevc in the laboratory frame \cite{btohh}.  
The probability of identifying a pion as a charged kaon 
is less than 2\%.
An electromagnetic calorimeter made of cesium iodide crystals 
provides energy measurements for electrons and photons, 
and an instrumented flux return detector identifies muons \cite{tautolll}.
For momenta above 1\gevc in the laboratory frame, 
electrons and muons are identified with efficiencies of approximately 
92\% and 70\%, respectively.  
Based on an integrated luminosity of 476\invfb, 
the data sample contains approximately 875 million \mtau leptons.

Simulated event samples are used to estimate the purity of the data sample.  
The production of \mtau pairs is simulated with the KK2F Monte Carlo (MC)
event generator \cite{kk}.  
Subsequent decays of the $\tau$ lepton, 
continuum \qqbar events (where $q=u,d,s,c$), 
and final-state radiative effects are modeled with Tauola \cite{tauola}, 
JETSET \cite{jetset}, and PHOTOS \cite{photos}, respectively.  
Passage of the particles through the detector is simulated by Geant4 \cite{geant}.

The $\tau$ pair is produced back-to-back in the \epem CM frame.
As a result, the decay products of the two $\tau$ leptons can be separated
from each other by dividing the event into two hemispheres -- 
the ``signal'' hemisphere and the ``tag'' hemisphere -- using the 
event thrust axis \cite{thrust}.  
The event thrust axis is calculated using 
all charged particles and all photon candidates in the entire event.
We select events with one prompt track and a \KS\to\pip\pim candidate 
reconstructed in the signal hemisphere, 
and exactly one oppositely charged prompt track in the tag hemisphere.  
A prompt track is defined to be a track with 
its point of closest approach to the beam spot being less than 1.5\cm  
in the plane transverse to the \en beam axis and 
less than 2.5\cm in the direction of the \en beam axis.  
Furthermore, if a pair of tracks is consistent with 
coming from a \KS or $\Lambda$ decay, or from a \g conversion 
after a mass cut and a displaced vertex cut, 
neither track can be a prompt track.  
The components of momentum transverse to the \en beam axis for 
each of these two prompt tracks must be greater 
than 0.1\gevc in the laboratory frame.  
The event is rejected if the prompt track in the signal hemisphere 
is identified to be coming from a charged kaon.
A \KS candidate is defined as a pair of oppositely charged pion candidates 
with invariant mass between 0.488 and 0.508\gevcc; 
furthermore, the distance between the beam spot and the \pip\pim vertex 
must be at least three times its uncertainty 
(the \pip\pim will be referred to as the ``\KS candidate daughters'').  
To reduce backgrounds from non-\mtau-pair events, 
we require that the momentum of the charged particle in the tag hemisphere 
be less than 4\gevc in the CM frame and 
be identified as an electron ($e$-tag) or a muon ($\mu$-tag).  
To reduce backgrounds from Bhabha, \mup\!\!\mun, and \qqbar events, 
we require the magnitude of the event thrust to be between 0.92 and 0.99.

\renewcommand\arraystretch{1.25}
\begin{figure}[ht]
\begin{center}
\includegraphics[width=8.5cm]{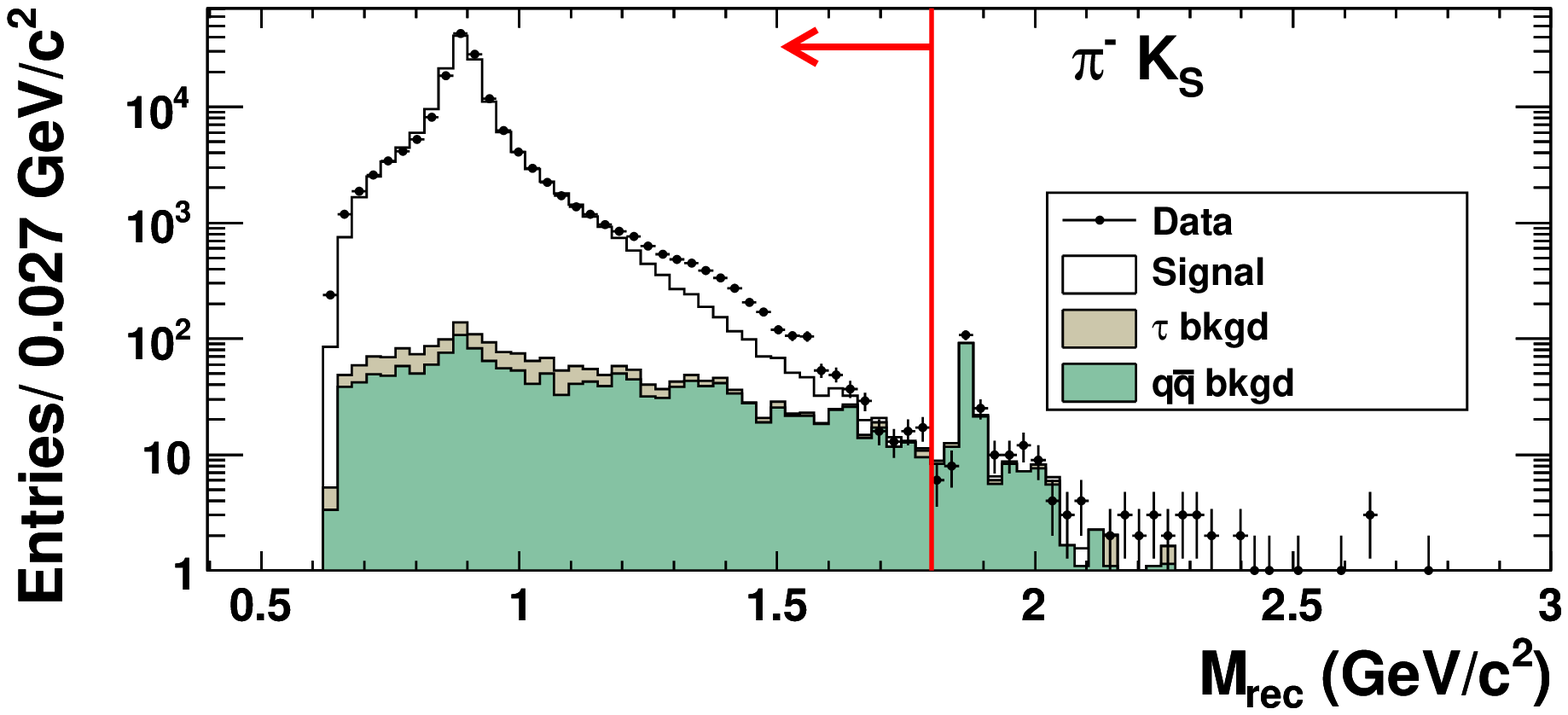}
\includegraphics[width=8.5cm]{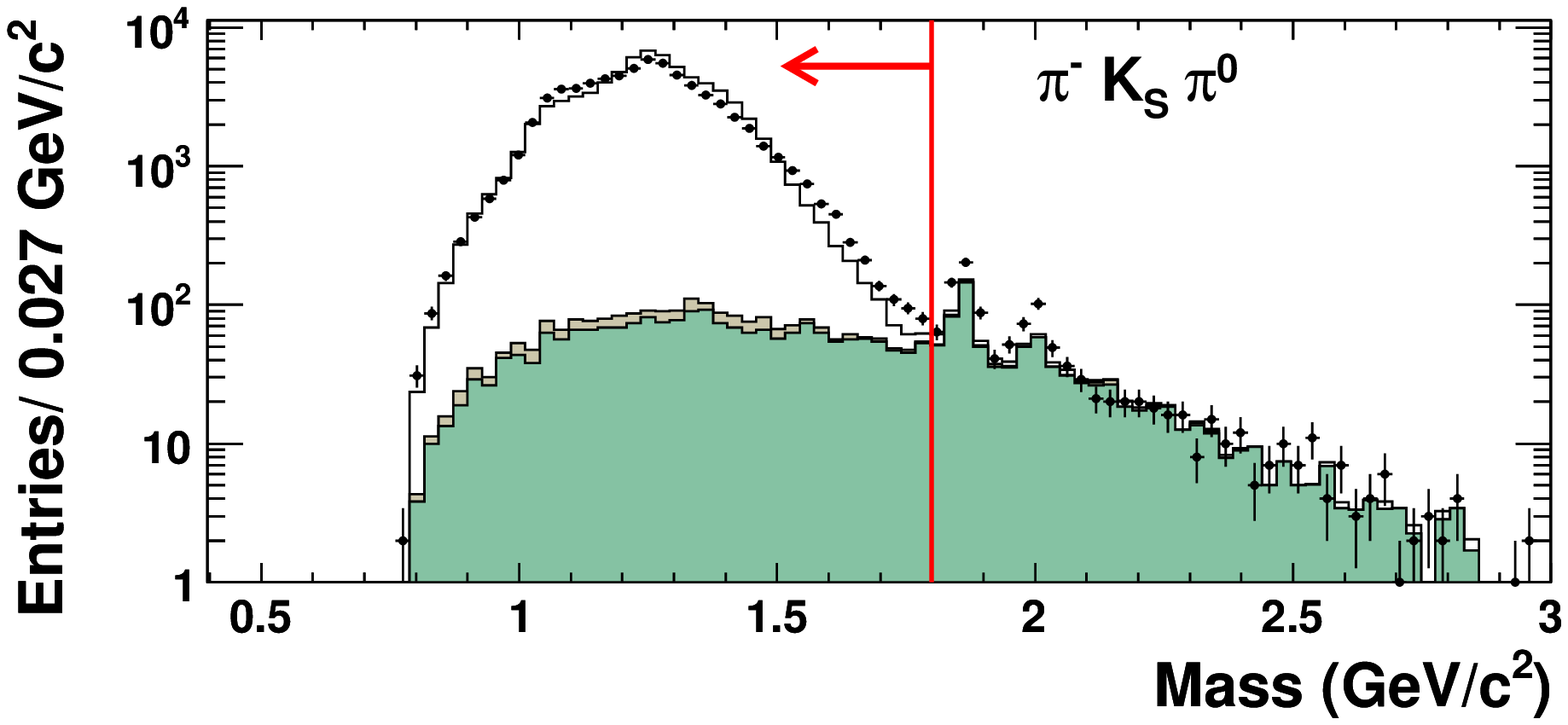} 
\includegraphics[width=8.5cm]{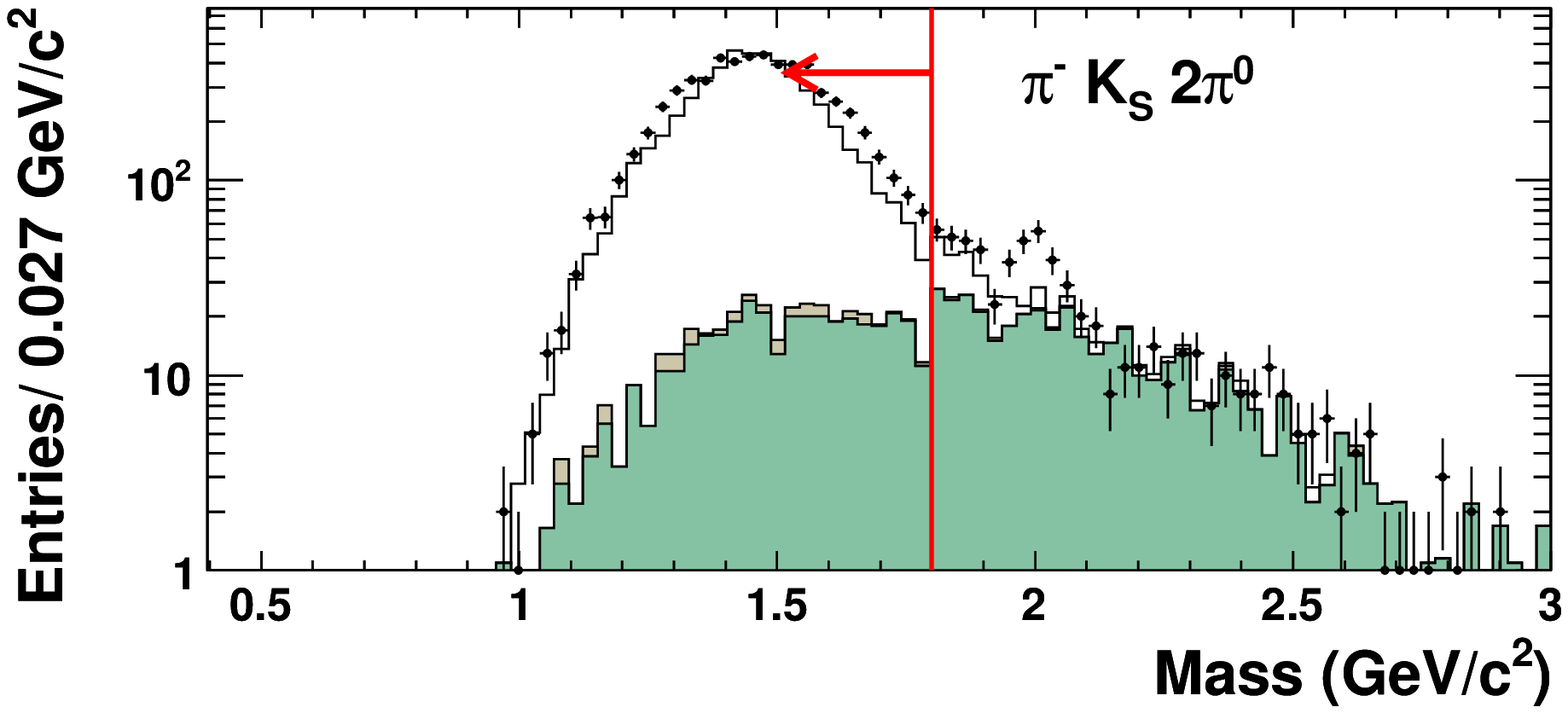} 
\caption{Invariant-mass distributions
for the combined $e$-tag and $\mu$-tag samples.
The label in each plot indicates the reconstructed decay mode 
(including the charge conjugate mode).
Points with error bars represent data whereas 
the histograms represent the simulated sample.
The histogram labeled as ``Signal'' includes the \taupksgpiz,
residual \taukksgpiz, and \taupkzkz modes.    
All selection criteria (including the likelihood ratio requirement), 
except the invariant mass ($M_{\rm rec}$) criterion, 
have been applied.  
The vertical lines and arrows indicate the 
$M_{\rm rec} < 1.8\gevcc$ selection criterion.
\label{fig:hksmass}}
\end{center}
\end{figure}

Backgrounds from \qqbar events are further reduced by 
rejecting events in which 
the invariant mass $M_{\rm rec}$ of the charged particle 
(assumed to be a pion), the \KS candidate, and up to three \piz candidates, 
all in the signal hemisphere, 
is greater than 1.8\gevcc (see Fig.~\ref{fig:hksmass}).  
If more than three \piz candidates are reconstructed 
in the signal hemisphere, 
the three with invariant masses closest to the \piz mass \cite{pdg} 
are included in the calculation of $M_{\rm rec}$ and 
the rest are ignored.
The \piz candidates are constructed from two clusters of energy deposits
in the electromagnetic calorimeter that have no associated tracks 
(``neutral clusters'').  
The energy of each cluster is required to be greater than 30\mev 
in the laboratory frame,  
and the invariant mass of the two clusters must be 
between 0.115\gevcc and 0.150\gevcc.  
The number of events in the \taupksthrpiz mode is small
and the corresponding invariant mass plot is not included 
in Fig.~\ref{fig:hksmass}.

The imperfect agreement between the $M_{\rm rec}$ distributions 
in the data and MC simulation, seen in Fig.~\ref{fig:hksmass}, 
is attributed to strange resonances that are not included 
in the simulation.
The impact of the modeling of the  $\tau$ decay modes in the MC simulation 
on the decay-rate asymmetry is found to be small and is included in 
the systematic uncertainties.

\begin{figure}[ht]
\centering
\includegraphics[width=8.5cm]{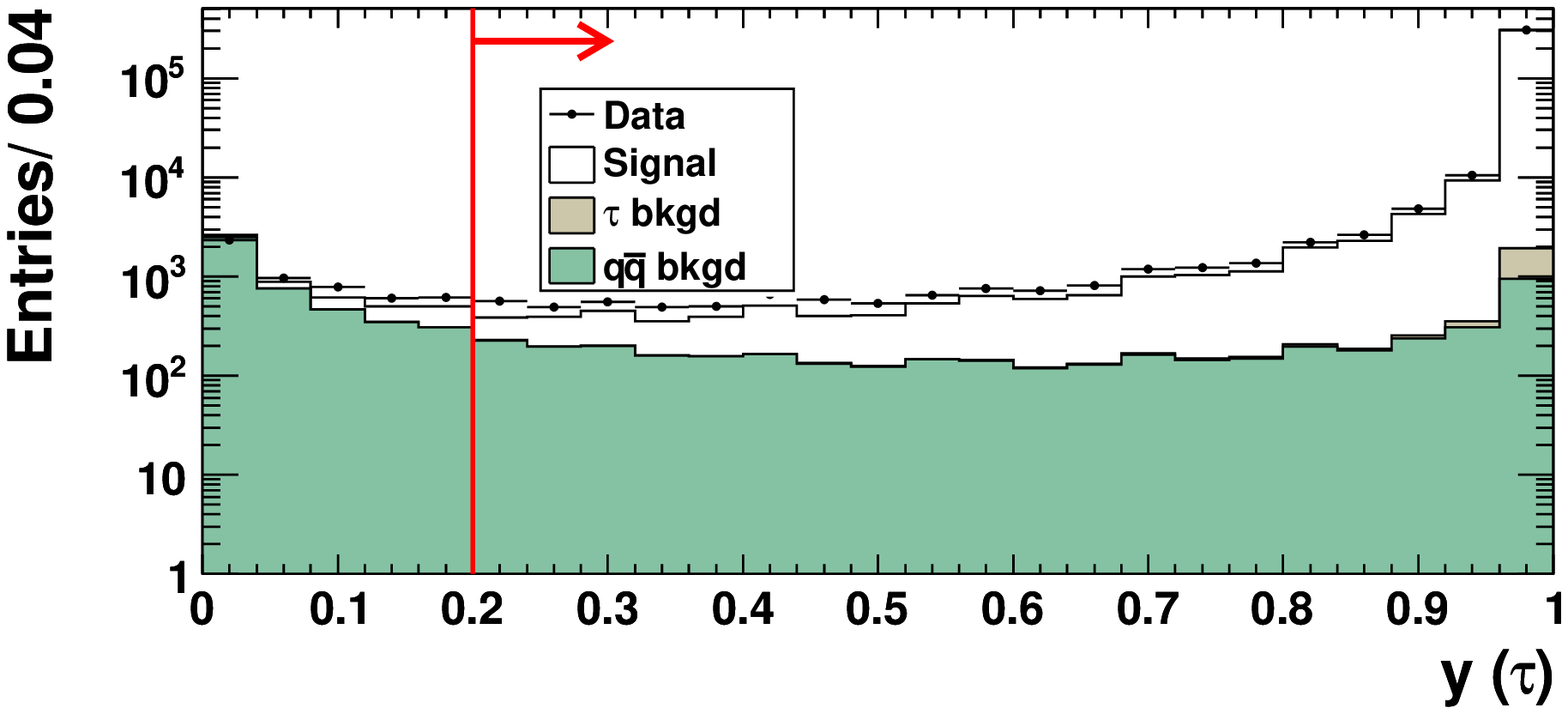} 
\includegraphics[width=8.5cm]{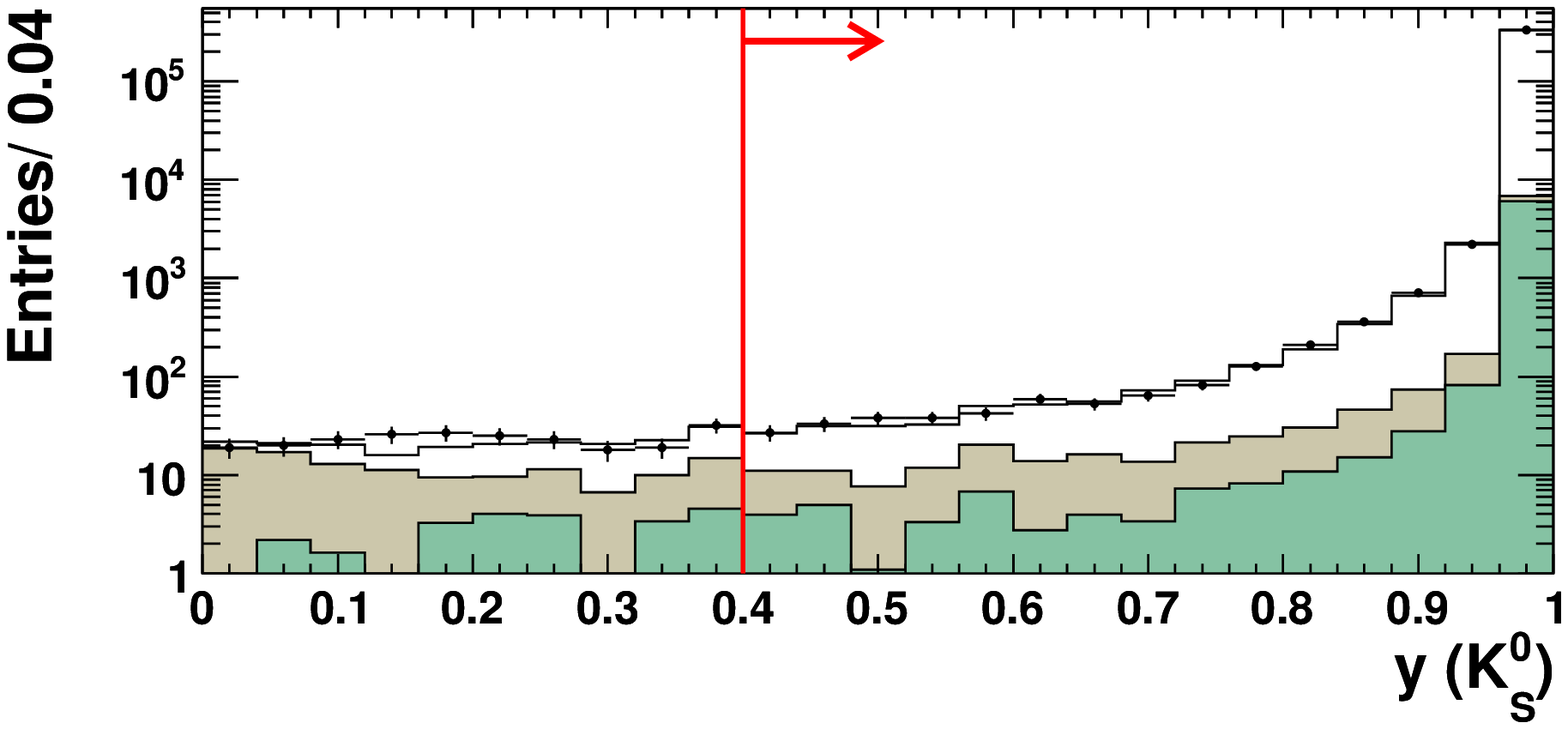} 
\caption{The likelihood ratio $y (\tau)$ used 
to distinguish $\tau$ events from \qqbar events 
(top plot) and the likelihood ratio $y (\KS)$ used to select $\tau$ decays with a 
$\KS \rightarrow \pip\pim$ (bottom plot).  
All selection cuts, except the plotted likelihood ratio requirement, 
have been applied.  
Points with error bars represent data while 
histograms correspond to simulated events.  
The histogram labeled as ``Signal'' includes the \taupksgpiz,
residual \taukksgpiz, and \taupkzkz modes. 
The vertical lines indicate the selection criteria.
 \label{fig:like}}
\end{figure}

A likelihood ratio $y (\tau)$ is used to distinguish $\tau$-pair events 
from \qqbar events, 
and a second likelihood ratio $y (\KS)$ is used to reduce 
the background in the sample of $\KS \rightarrow \pip\pim$ candidates.
The likelihood ratio $y_{i}\left(\vec{x_i}\right)$, 
where $i$ refers to $\tau$ or \KS, 
is defined as 
$y_{i}(\vec{x_i}) \equiv 
\mathcal{L}_i^s(\vec{x_i}) / 
(\mathcal{L}_i^s(\vec{x_i}) + w\mathcal{L}_i^b(\vec{x_i}))$
where $w$ is the background-to-signal ratio 
estimated from the MC simulation, 
$\mathcal{L}_i^s$ ($\mathcal{L}_i^b$) is the likelihood function 
for signal (background) events, and 
$\vec{x}_i$ is the set of variables used for likelihood $i$.  
Each likelihood function is a product of one-dimensional 
probability distribution functions of the variables $\vec x_i$ 
obtained from the MC simulation.
For $y (\tau)$, the variables $\vec x_i$ are 
the visible energy (sum of the energies associated with 
all neutral calorimeter clusters 
and tracks in the event),
the number of neutral clusters in the tag hemisphere,
the number of neutral clusters in the signal hemisphere,
the magnitude of the thrust, and 
the component of the total momentum of the event 
transverse to the \en beam axis 
(calculated from all tracks and neutral clusters 
in both hemispheres).  
The variables used to construct $y (\KS)$ are 
the distance from the beam spot  
to the decay vertex of the 
\KS candidate in the plane transverse to the \en beam axis, 
the invariant mass of the \KS candidate daughters,
the magnitude of the \KS momentum, and 
the cosine of the polar angle of the \KS candidate.  
The polar angle is the angle between 
the \KS trajectory and the \en beam axis.  
The cosine of the polar angle discriminates 
low-angle photon conversions from genuine \KS candidates.  
All kinematic quantities 
used in the construction of the two likelihood ratios, 
except for thrust, 
are determined in the laboratory frame.
Events are selected if $y (\tau) > 0.2$ and $y (\KS) > 0.4$ 
(see Fig.~\ref{fig:like}), 
in order to minimize the contamination from background events while 
maintaining a high selection efficiency.

After all selection criteria are applied, 
a total of \ndate\ (\ndatm) candidates are obtained 
in the $e$-tag ($\mu$-tag) sample, 
of which there are \ndatpe\ (\ndatpm) in the \taum\ sample
and \ndatne\ (\ndatnm) in the \taup\ sample.


The sample contains events from two \mtau decay modes,
\taukksgpiz and \taupkzkz, that also have \KS mesons in the final state.
The decay \taupkzkz satisfies the selection criteria if one of the 
neutral kaons decays into \pip\!\!\pim and the other neutral kaon decays 
into 2\piz or appears as a \KL meson.
  
The selected candidate sample also contains a small background component 
from $\tau$ decays not containing a \KS in the final state, 
as well as continuum \qqbar ($u$, $d$, $s$ and $c$-quark) events.
There is no background from $B\overline{B}$ events.

The numbers of background events of each type are estimated from the 
MC simulation.  
The accuracy of the background estimation is evaluated by measuring 
the ratios of data to simulated event yields in the region 
$y (\tau) < 0.1$ and $y (\KS) < 0.1$.  
A correction factor is then applied to the background yield 
estimated from the Monte Carlo simulation in this region.  
The correction factors are determined to be 
\corrtaue$\pm$\ecorrtaue\ (\corrtaum$\pm$\ecorrtaum) 
for the \qqbar background and 
\corrkse$\pm$\ecorrkse\ (\corrksm$\pm$\ecorrksm) 
for the non-\KS \mtau background  in the $e$-tag ($\mu$-tag) samples, respectively.
The total numbers of background events are then estimated to be
\nbkgpe$\pm$\ebkgpe\ (\nbkgpm$\pm$\ebkgpm) 
for \taum\ decays and 
\nbkgne$\pm$\ebkgne\ (\nbkgnm$\pm$\ebkgnm) 
for \taup\ decays in the $e$-tag ($\mu$-tag) samples, 
where all selection criteria (including the requirements on the two 
likelihood ratios) are applied.
The uncertainties include the statistical uncertainties from the sizes 
of the Monte Carlo samples and the uncertainties of the correction factors.
The composition of the sample is given Table \ref{table:results}.

After the subtraction of background composed of 
\qqbar and non-\KS $\tau$ decays, 
the decay-rate asymmetry is measured to be
$\aeraw$ for the $e$-tag sample and 
$\amraw$ for the $\mu$-tag sample,
where the errors are statistical.  

\renewcommand\arraystretch{1.25}
\begin{table}[ht]
\caption{Breakdown of the sample after all selection criteria 
have been applied.  
The errors of the decay modes with \KS are dominated by the 
uncertainties in the branching fractions.
The background from other \mtau decays and \epem\to\qqbar background 
are estimated using the data and MC simulation samples.
\label{table:results}}
\begin{center}
\begin{tabular}{lcc} 
\hline \hline
Source                         & \multicolumn{2}{c}{Fractions (\%) } \\
                               & \hspace{0.5cm} $e$-tag \hspace{0.5cm}
                               & \hspace{0.5cm} $\mu$-tag \hspace{0.5cm} \\ 
\hline 
$\tau^- \rightarrow \pim \,\KS (\geq 0\piz) \, \nut$  & $78.7 \pm 4.0$  & $78.4 \pm 4.0$  \\
$\tau^- \rightarrow \Km  \,\KS (\geq 0\piz) \, \nut$  & $4.2 \pm 0.3$   & $4.1 \pm 0.3$   \\
$\tau^- \rightarrow \pim \,\Kz \Kzb \, \nut$          & $15.7 \pm 3.7$  & $15.9 \pm 3.7$  \\
Other background      & $1.40 \pm 0.06$ & $1.55 \pm 0.07$  \\
\hline \hline
\end{tabular}
\end{center}
\end{table}

A control sample of $\tau^- \rightarrow h^- h^- h^+ (\geq 0\piz) \, \nut$  
(excluding $\KS \rightarrow \pip \pim$ decays) in both data and MC simulation, 
where $h^-$ ($h^+$) represents a negatively (positively) charged hadron,
is used to confirm that no significant decay-rate asymmetry
is induced by the \babar\ detector or the selection criteria.  
The control sample is selected by requiring that all charged tracks 
be prompt tracks, which 
suppresses \KS\ contamination due to its displaced decay vertex.  
The asymmetries measured in the simulated and data control samples agree 
to within the experimental uncertainties of the measurements, which are 
$\deterre\%$ for the $e$-tag and $\deterrm\%$ for the $\mu$-tag, and 
include both statistical and systematic components.  
These errors are taken as systematic uncertainties on the signal asymmetry 
(see Table~\ref{table:final}).  

\renewcommand\arraystretch{1.25}
\begin{table}[ht]
\caption{Summary of systematic uncertainties in the decay-rate asymmetries.
\label{table:final}}
\begin{tabular}{lcc} \hline\hline
                               & \hspace{0.3cm} $e$-tag   \hspace{0.3cm}     
                               & \hspace{0.3cm} $\mu$-tag \hspace{0.3cm}   
\\ \hline
Detector and selection bias    & \deterre\%          & \deterrm\%    \\
Background subtraction         & \bkgerre\%          & \bkgerrm\%    \\
\Kz/\Kzb interaction           & \kzerrnew\%         & \kzerrnew\%     \\ 
Total                          & \totsyse\%          & \totsysm\%    \\ 
\hline \hline
\end{tabular}
\end{table}

Additional studies show no evidence for any charge-dependent biases in the
selection criteria.
We find no decay-rate asymmetry in the MC sample of \taupksgpiz decays
(no \CP violation is modeled in the simulation) where the error on the 
decay-rate asymmetries is 0.14\% for the $e$-tag and 0.17\% for the $\mu$-tag events.
We vary the selection criteria around their nominal values,
and no significant changes in the asymmetry are observed.
The decay-rate asymmetry of the background events was studied by examining
the events rejected by the likelihood ratio criteria 
and was found to be consistent with zero  
for both data and MC simulation.  

A recent paper~\cite{ko} suggests that the decay-rate asymmetry
will be modified due to the different nuclear-interaction cross sections 
of the \Kz and \Kzb mesons with the material in the detector. 
This effect is not included in the MC simulation.
A correction to the asymmetry accounting for this effect is calculated on an 
event-by-event basis using the momentum and polar angle of the \KS 
candidate together with the nuclear-interaction cross sections 
for neutral kaons, which are related by isospin symmetry to the $K^\pm$ 
nucleon cross sections \cite{pdg}.
The kaon-nucleus cross sections are determined by using the kaon-nucleon
cross sections and including a nuclear screening factor of 
$A^{0.76}$, where $A$ is the atomic weight \cite{ko}.
The correction, which is subtracted from the measured asymmetry,
is found to be 
$\ad$
for both the $e$-tag and the $\mu$-tag samples.
The error includes the statistical uncertainty in the MC simulation, 
the uncertainties in the kaon-nucleon cross sections \cite{pdg}, 
and an uncertainty due to the assumption of isospin invariance.  
The latter effect is taken to be 5\% by observing that isospin symmetry 
in pion-nucleon cross sections holds to within a few percent.
The error on the exponent of the atomic weight of the nuclear screening 
factor is 0.003 \cite{ko} and its contribution to the uncertainty
in the asymmetry correction is negligible.

The measured decay-rate asymmetries 
(after correcting for the difference in neutral kaon nuclear interactions)
are 
$\aed$  for the $e$-tag sample and
$\amd$  for the $\mu$-tag sample,
where the first error is statistical and the second is systematic.
The systematic uncertainties of the $e$-tag and $\mu$-tag results 
are almost completely uncorrelated.  
The small correlations in the systematic uncertainties for the two samples 
are ignored when the average is computed.
The weighted average of the two decay-rate asymmetries is 
$\amean$.

The asymmetry measured at this stage 
still includes other $\tau$ decays with \KS in the final state.  
Specifically, the decay-rate asymmetry is diluted 
due to \taukks and \taupkzkz decays.  
The measured asymmetry $\mathcal{A}$ is related to the signal asymmetry 
$A_1$ and the remaining background asymmetries 
$A_2$ and $A_3$ by:
\begin{eqnarray*}
\mathcal{A} & = & \frac{f_1 A_1 + f_2 A_2 + f_3 A_3}
                       {f_1 + f_2 + f_3} \\ 
            & = & \left(\frac{f_1 - f_2}{f_1 + f_2 + f_3}\right)\asy
\end{eqnarray*}
\noindent where $f_1$, $f_2$, and $f_3$ are, respectively, 
the fractions of \taupksgpiz, \taukksgpiz, and \taupkzkz 
in the total selected sample, shown in Table~\ref{table:results}.  
Within the SM, 
\(A_1 = -A_2  \) 
because the \KS in \taupksgpiz is produced via a \Kzb, 
whereas the \KS in \taukksgpiz is produced via a \Kz.  
Furthermore, \(A_3 = 0\) in the SM because the asymmetries 
due to the \Kz and \Kzb will cancel each other.
Using the relations between $A_1$, $A_2$, and $A_3$,
we can compare our result with the theoretical prediction by 
dividing the measured decay-rate asymmetry of
$\mathcal{A} = \amean$
by  $(f_1 - f_2)/(f_1 + f_2 + f_3) = 0.75 \pm 0.04$
(the correction is identical for the $e$-tag and $\mu$-tag samples).
The uncertainty on the correction includes the statistical uncertainty 
and uncertainties in the branching fractions.  
Finally, the decay-rate asymmetry for the \taupksgpiz decay 
for the combined $e$-tag and $\mu$-tag sample
is calculated to be 
$\asy = \afinal$.

\begin{figure}[ht]
\begin{center}
\mbox{\epsfig{file=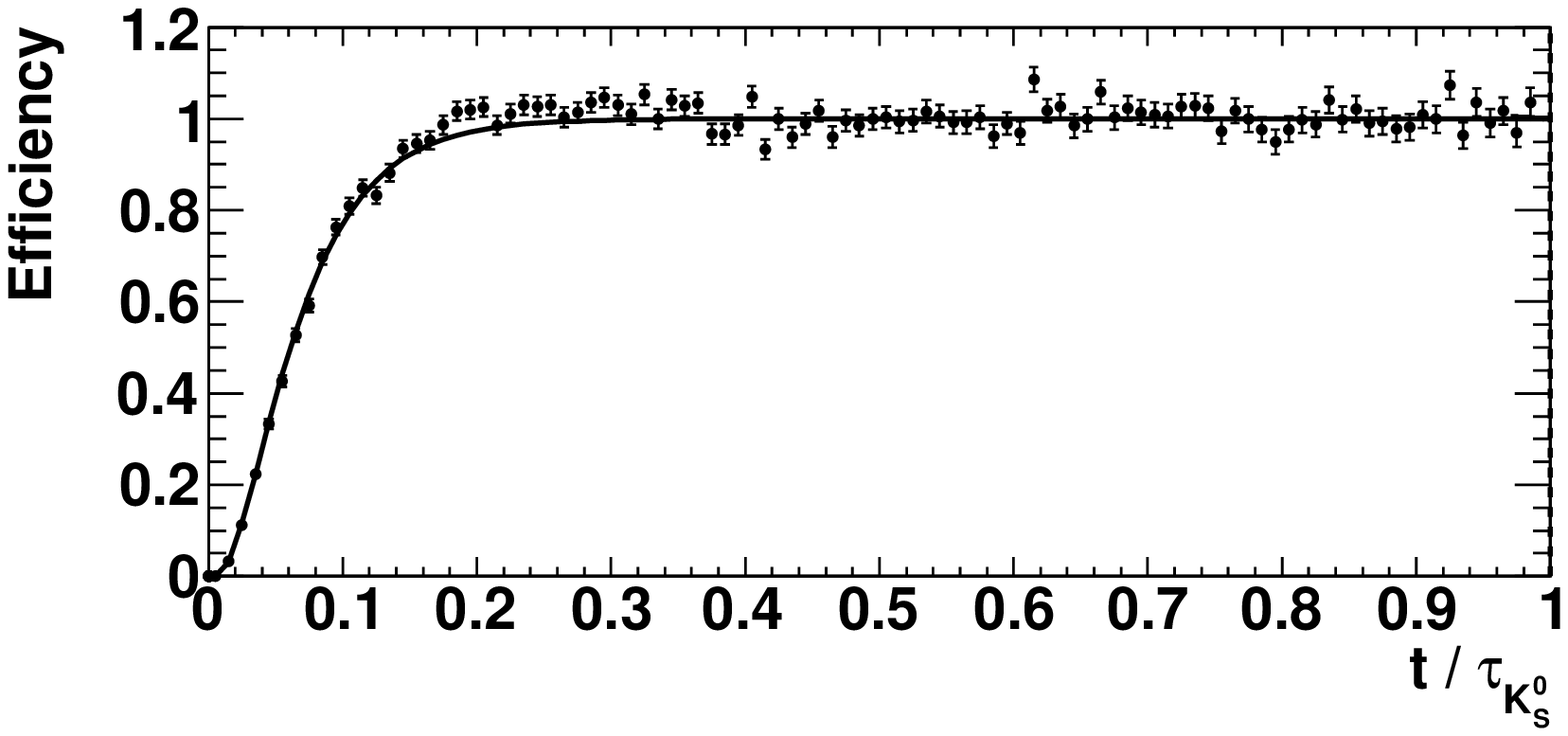,height=4.25cm}}
\mbox{\epsfig{file=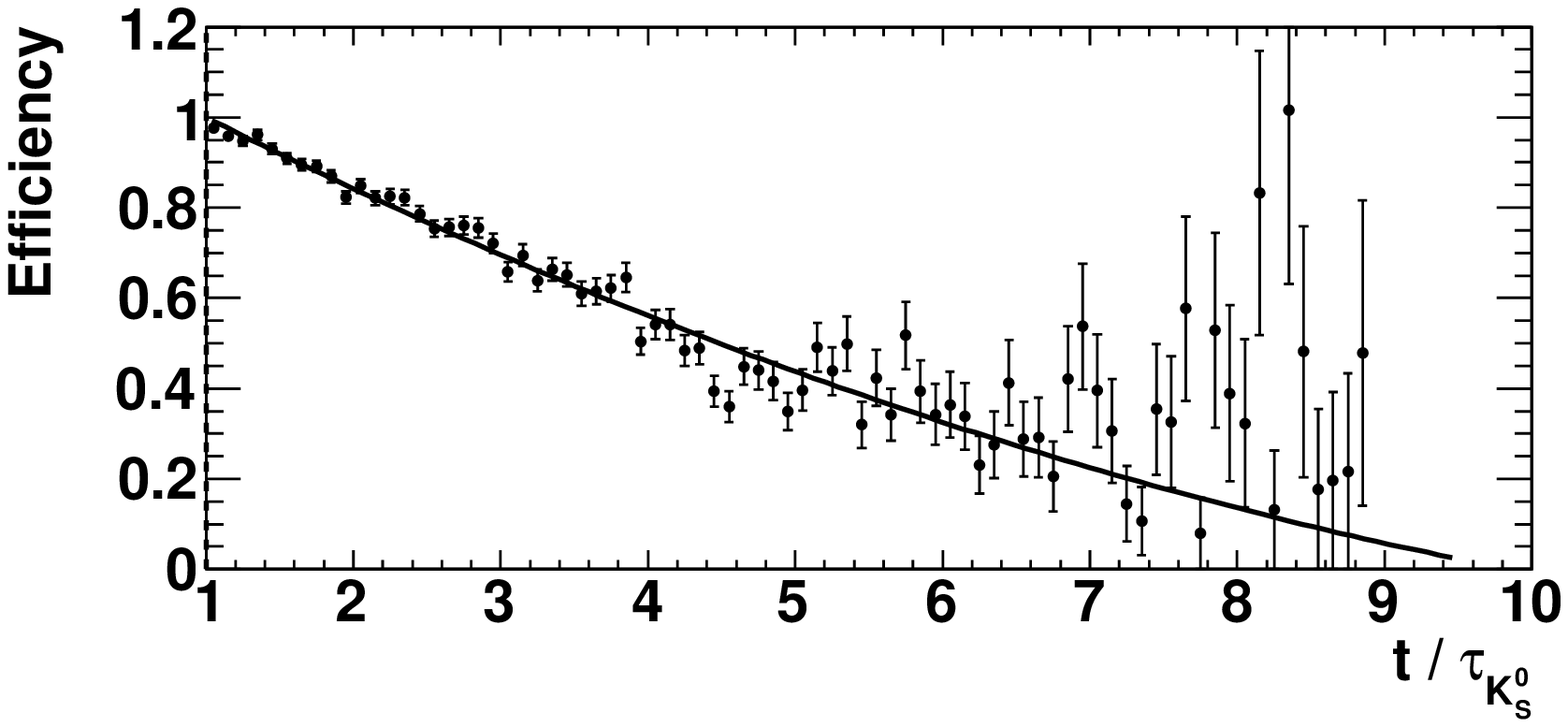,height=4.25cm}}
\caption{The relative selection efficiency as a function of $\tks$
obtained from the Monte Carlo sample.
The top plot shows the region $0 < \tks <1$ and the bottom plot
the region $1 < \tks < 8$.
The solid line is the fit to the points in the displayed region.
The relative efficiency is normalized to be unity for the region
$0.25 < \tks < 1.0$.
\label{fig:efficiency}}
\end{center}
\end{figure}

As pointed out by Grossman and Nir, the predicted decay-rate asymmetry is 
affected by the $\KS \rightarrow \pip \pim$  decay time dependence of 
the event selection efficiency \cite{grossman}.
Figure~\ref{fig:efficiency} shows the relative selection efficiency,
defined as the selection efficiency normalized to unity 
in the range $0.25 < \tks < 1.0$.
In the  $0< \tks < 1$ region, the relative  efficiency is parametrized 
with the function $\left( 1 - A e^{-B (t - t_0)} \right)^{-2}$,
where $A$, $B$, and $t_0$ are constants.
In the  $1 < \tks < 8$ region, the relative efficiency 
is parametrized by a second-order polynomial.
Both functions are constrained to unity at $\tks=1$.
We use this parametrization in
Eq.~(13) of the Grossman and Nir 
paper \cite{grossman} to obtain a multiplicative correction factor of  
$1.08 \pm 0.01$ for the decay-rate asymmetry, where the error is 
due to the uncertainty in the relative selection efficiency.
After applying the correction factor, the 
SM decay-rate asymmetry is predicted to be ($0.36 \pm 0.01$)\%.

In conclusion, we have performed a search 
for \CP violation using the \taupksgpiz decay mode.
The decay-rate asymmetry is measured for the first time and is found to be 
$\afinal$.
The measurement is $\nstd$ standard deviations from 
the SM prediction of $(0.36 \pm 0.01)\%$.

The authors thank Y. Grossman and Y. Nir for their useful suggestions.
We are grateful for the excellent luminosity and machine conditions
provided by our \pep2\ colleagues, 
and for the substantial dedicated effort from
the computing organizations that support \babar.
The collaborating institutions wish to thank 
SLAC for its support and kind hospitality. 
This work is supported by
DOE
and NSF (USA),
NSERC (Canada),
CEA and
CNRS-IN2P3
(France),
BMBF and DFG
(Germany),
INFN (Italy),
FOM (The Netherlands),
NFR (Norway),
MES (Russia),
MICIIN (Spain),
STFC (United Kingdom). 
Individuals have received support from the
Marie Curie EIF (European Union),
the A.~P.~Sloan Foundation (USA)
and the Binational Science Foundation (USA-Israel).


\end{document}